\documentstyle[epsfig,12pt]{article}
\newcommand\beq{\begin{equation}}
\newcommand\eeq{\end{equation}}
\newcommand\bea{\begin{eqnarray}}
\newcommand\eea{\end{eqnarray}}
\newcommand\blankline{{\vskip 1cm}}
\newcommand\C{{\cal C}}
\newcommand\World{\mbox{\sl World}}
\newcommand\Empty{\mbox{\sl Empty}}
\newcommand\Past{\mbox{\sl Past}}
\newcommand\History{\mbox{\sl History}}
\newcommand\mapdown[1]{\Big\downarrow
        \rlap{$\vcenter{\hbox{$\scriptstyle#1$}}$}}
 
\newcommand\mapright[1]{\smash{
                \mathop{\mbox{\large{$\longrightarrow$}}}\limits^{#1}}}

\newcommand\mapdots[1]{\smash{
                \mathop{\mbox{\large{$\cdots$}}}\limits^{#1}}}
 
\newcommand\bundle[3]{\begin{array}[t]{c}
                {#1}\\ \mapdown{#2}\\ {#3}\end{array}}
 
\newcommand\bundlemap[2]{\begin{array}[t]{c}
     \mapright{#1}\\ \phantom{\mapdown{}}\\\mapright{#2}\\\end{array}}

\newcommand\bundledots[2]{\begin{array}[t]{c}
     \mapdots{#1}\\ \phantom{\mapdown{}}\\\mapdots{#2}\\\end{array}}
\begin{document} 
\begin{flushright}
{\small  CGPG-98/11-2}
\end{flushright}
\centerline{\Large\bf The internal description of a causal set:}
\vskip 0.2cm
\centerline{\Large\bf What the universe looks like from the inside}
\vspace{0.5in}
\rm
\centerline{Fotini Markopoulou}
\blankline
\centerline{\it  Center for Gravitational Physics and Geometry}
\centerline{\it Department of Physics}
\centerline {\it The Pennsylvania State University}
\centerline{\it University Park, PA, USA 16802 }
\centerline{fotini@phys.psu.edu}
\vfill
\centerline{November 13, 1998}
\vfill
\centerline{\bf Abstract}
\blankline
{\small{We describe an algebraic way to code the causal information
of a discrete spacetime. 
The causal set $\C$ is transformed to a description in 
terms of the causal pasts  of the events in $\C$.
This is done by an evolving set, 
a functor which to each event of $\C$ assigns its causal 
past. Evolving sets obey a Heyting algebra which is characterised 
by a non-standard notion of complement. 
Conclusions about the causal structure of the causal set
can be drawn
by calculating the complement of the evolving set.
A causal quantum theory can be based on the 
quantum version of evolving sets, which we briefly discuss.}}

\vfill
 
\section{Introduction}

In general, an entire spacetime, or an entire spatial slice in canonical
gravity, can only be seen by an observer either in the infinite 
future or outside the universe. This is unphysical, so in the 
present paper we look for an alternative description of the 
causal structure of the spacetime that codes what an observer
inside the universe can observe.

Another motivation for this work is to develop the proposal 
advocated in \cite{fmls1}, that causality
persists at the planck scale.
Before one can argue whether or not this is plausible,
there is the question of how to describe the causal structure. 
A spacetime metric is totally unsuitable since it is classical;
when quantized it ``fluctuates'', leading to confusion in 
any argument, either for, or against, planckian causality.

An alternative to the Lorenzian metric which has been explored,
for example, in \cite{sorkin} is a causal set. 
This is the discrete equivalent of 
a Lorentzian spacetime, the set of 
events in a discrete spacetime, partially ordered by the 
causal relations. Because the causal 
set can be defined prior to a spacetime manifold, it 
has been used in quantum gravity approaches that 
hold that a manifold is a classical concept, 
only found in a suitable classical limit of the 
planckian theory. Incorporating a causal set,
as for example in \cite{fmls1,dual,fmls2}, 
makes a theory ``in principle'' causal.  However, the causal set 
itself is simply a very large collection of causal relations. 
If causality can instead be coded in an algebra, it may be
possible to represent it directly in a quantum theory.

In order to address both of the above problems, we wish to
introduce a transformation 
from the causal set $\C$ to the set of causal pasts of 
each event in $\C$.
One possibility is to simply replace each event with its causal past.
This has some advantages but it makes no progress towards a 
better coding of the causal structure. The maps between the causal
pasts will be given exactly as in the causal set case.
Therefore, we instead
choose to work with {\it evolving sets}. An evolving set 
is a functor from the causal set to the category of 
sets. It is a generalization
of an ordinary set that varies over the events in $\C$. It  gives
the causal past of each event but also contains the causal 
structure of $\C$ in an intrinsic fashion.

Evolving sets can be generalised to causal quantum theories,
as we will discuss in this paper.  Evolving sets satisfy a
particular algebra, called a Heyting algebra, with operations
which reflect the underlying causal set. 
In particular, the Heyting algebra complement is 
a measure of the non-triviality 
of the causal structure of $\C$. This algebra 
corresponds to a type of non-standard logic whose development, 
historically, is closely related to the passage of time.

In short, the evolving set of causal pasts
contains the same information about the causal structure as 
the causal set, but it this information can be given
in terms of algebraic relations. 
An algebraic way to 
express time evolution--- as opposed to time given
geometrically as an extra dimension---is likely to be advantageous in
addressing quantum gravity issues. 

In broader terms, the use of varying sets,
of which the evolving ones which we study  here are a particular
case (those that vary over 
a causal set), applies to several occasions in physics when we 
need to make explicit  
the conditions under which a physical statement is made. In the present
work, and generally in a causal quantum theory, we need to make 
explicit the time (event) dependence. Isham \cite{cji} and Isham and 
Butterfield \cite{IB} who introduced varying sets in physics have used them
to specify consistent sets of histories and the different 
levels of coarse-graining. 
 
The outline of the paper is the following.
In section \ref{causal}, we give the definition of a causal set and
explain when we regard the causal structure
of $\C$ to be trivial. In section \ref{outline}, we give a preview of
the final construction of the past-sets over $\C$ to indicate what 
technical tools are needed and motivate
some of the rather technical sections that follow. Since
the evolving set of causal pasts is  a generalisation of an ordinary set, in 
section \ref{sets} we review material from set theory. In particular, 
we examine the definition of a subset of some set in standard set theory.
Then, we give the definition of a subset in categorial terms, since this
is the form which we will generalise. The mathematical side of this 
material can be found in the 
standard mathematical literature on topos theory, for example in 
Mac Lane and Moerdijk \cite{McLM}, and we mainly concentrate on its 
physical interpretation.

In section \ref{past}, 
we construct the evolving set over a general causal set. 
In section \ref{newtonian}, 
we apply this definition to the example of discrete 
Newtonian histories, a particularly simple case of a causal set.
In section \ref{complement}, we define the complement of a causal 
set and show that 
it is empty when the causal set is a lattice. The complement provides
an algebraic definition of event-dependent causal horizons and we 
examine the 
possibility of generalising this to global properties of the 
causal set, such as black holes and branchings.

The complement is one of the operations of the Heyting algebra which 
the evolving sets obey. We give the definitions of all the four 
operations in a Heyting algebra  in section \ref{algebra}, 
and we compare it to the standard Boolean algebra of ordinary sets.  

Just as the boolean algebra of set theory implies an underlying 
boolean logic, the Heyting algebra of evolving sets means 
that the underlying logic is intuitionistic. For completeness,
we translate the algebraic operations of Section \ref{algebra}
to logical ones in the Appendix. Finally, in section \ref{quantum},
we outline possible generalisations of the present
results to causal spin networks and the quantum theory.

\section{Review of causal sets}
\label{causal}
A causal set $\C$ is a discrete partially ordered set  with structure 
that is intended to  mirror that
of Lorentzian spacetime. Namely, for any pair of points $p$ and $q$ either
one is to the future of the other, say $p\leq q$, 
or they are causally unrelated. 
The ordering relation is antisymmetric,  if $p\leq q$ and 
$q\leq p$, we must conclude that $p=q$ since
timelike loops are not allowed.
That the causal set is discrete means that the cardinality of  the set  
$\{r\in\C:p\leq r\leq q\}$  for any pair $p\leq q$ is finite. 
We will also assume that the causal set has a finite number of 
elements. 

Two elements $p$ and $q$ in the causal set
have greatest lower bound (g.l.b.), $r$, 
when $r$ is an element in the causal set such that $r\leq p$ and 
$r\leq q$ and, for any other  element $z$,
$z\leq p$ and $z\leq q$ implies $z\leq r$. Similarly, $p$ and $q$ have a 
least upper bound (l.u.b.) when there is an element $t$ in the causal set
such that 
$p\leq t$ and $q\leq t$, and if there is some element $z$ later than 
both $p$ and $q$, then it must also be later than $t$. The existence of 
a l.u.b.\ for two elements means that they will eventually meet at
that common future event, while their g.l.b.\ is their last common
past event. 
 
A supremum for a partially ordered set, if it exists, 
is an element $s$ in the partially ordered set later than every other 
element, that is, 
$p\leq s$ for all $p\in\C$. Similarly, an infimum is an element 
$i$ in the partially ordered set before every other element, 
$i\leq p$ for all $p\in\C$. An infimum in the causal set
means a single first event for the whole universe, while a 
supremum is the single final event that all others lead to.

A {\it lattice} is a partially ordered set with  
a g.l.b.\ and a l.u.b.\  for every 
pair of elements $(p,q)$ and a supremum and infimum.
We are mainly interested in causal sets with a finite number of events. 
In this case, if there is a g.l.b.\ and a l.u.b.\ for 
each pair of events in the causal set, there is also
a supremum and an infimum. 
When a causal set is a lattice, we will say that it has trivial causal 
structure.  The reason for this is explained in section \ref{complement}.
 
We use the causal set to express time-dependence. The
``position'' of some event in the causal set
is determined by its causal relations to the rest of the 
events in the causal set. To emphasize the fact that we use an event
$p$ to specify a time instant relative to the rest of $\C$, 
we will often call it the {\it stage}  $p$.

\section{The evolving set of causal pasts: the basic idea}
\label{outline}

A classical observer at a given time instant can be placed
on the  corresponding  
event $p$ in the causal set without affecting the causal set.
He knows about all the events in the  past of $p$. 
In this minimal description, time passes when the observer moves to a 
later stage $q\geq p$ in the causal set. At $q$, 
 the set of events in 
his causal past is larger but still includes all the past
events of $p$.

If we are ``outside'' the causal set, what we see is 
a collection of such causal pasts, one for each event in $\C$.
It is the thesis of this paper, however, that 
being outside the causal set is unphysical. We instead 
care about the same situation as viewed from ``inside'' $\C$ by one of the
above observers. We want to know in what way the inside viewpoint is different
from the outside one and if it has some interesting structure.

We will show that the inside viewpoint is given by a functor from the 
causal set  
category to the category of sets. Upgrading the causal set to 
a category involves nothing more than calling the events ``objects'' 
and the ordering relation $p\leq q$ , when it exists, an ``arrow $p\rightarrow q$''.
A collection of objects and arrows forms a category when the composition 
of the arrows is associative and there is an identity arrow. 
This is certainly the case for the causal set. From $p\leq q\leq r\leq s$ 
we can conclude that $p\leq s$ independently of the order in which
we removed the 
mid-events (associativity) and  
$p\leq p$ for any $p\in\C$ (identity). 

The category of sets, {\bf Set}, which we need for the sets of causal
pasts, has sets for its objects and functions between sets for arrows. Composition of 
functions is associative, $(f\circ g)\circ h=f\circ(g\circ h)$, 
and for each set there is an identity map from the set to itself. 

Our task is the following: we need to go from the causal set $\C$ 
to  the events in the past of  each $p\in\C$. The former belong to the causal set,
while the latter are objects in ${\bf Set}$.
One can go from one category to 
another, while preserving the structure of the first,
by a  functor. So, we 
employ the functor $\Past$ from the causal set to 
the category of sets, 
\beq
\Past:\C\rightarrow {\bf Set}.
\eeq
It has components $\Past(p)$
for each $p\in\C$ which are the events that lie in the 
past of that $p$,
\beq
\Past(p)=\{r\in\C :r\leq p\}.
\eeq
They are sets.
That $\Past$ is a functor means not only that it spits out 
$\Past(p)$ for each $p$ but also that 
whenever $p\leq q$  it gives
\beq
\Past_{pq}:\Past(p)\rightarrow \Past(q)\label{eq:Pastpq}
\eeq
which takes the causal past of $p$ to the causal past of $q$.
$\Past$ preserves composition, 
\beq
\Past_{pr}=\Past_{qr}\circ \Past_{pq}\qquad\mbox{whenever}\qquad
p\leq q\leq r,
\label{eq:Pastpqr}
\eeq 
and $\Past_{pp}$ is the identity map,
\beq
\Past_{pp}:\Past(p)\rightarrow \Past(p).
\label{eq:Pastpp}
\eeq

Clearly, when $p\leq q$, the past of $q$ contains the past of $p$
and the map $\Past_{pq}$ of equation (\ref{eq:Pastpq}) is 
really just set inclusion,
\beq
\Past_{pq}: \Past(p)\subseteq \Past(q).
\eeq
(The reason we bother with properties (\ref{eq:Pastpqr}) and
(\ref{eq:Pastpp}) which are obvious when $\Past_{pq}$ is set
inclusion is that we are interested in later generalising
the functor $\Past$ and assign, for example, open spin network
states  to each $p$, 
or turn $\Past_{pq}$ into a dynamical evolution arrow.)

The main role of $\Past$ is to transform events in $\C$ into sets (of events).
In fact, functors from a causal set to ${\bf Set}$,
like $\Past$, themselves form a category. 
These functors are not far from being sets, which is why they can be thought of 
as {\it evolving}, or {\it varying} {\it sets}, a generalisation of ordinary sets which can be found, for example, in \cite{McLM}. 
We will, therefore,  understand $\Past$ by generalising
standard set theory. We start by giving a review of the 
relevant material for sets.

\section{Sets and subsets}
\label{sets}

We start with a review of material from set theory.  Consider a 
set $X$ and a subset of it, $A$. How do we list those 
elements $x\in X$ which are also contained in the subset $A$?
We use the {\it characteristic function} for $A$, 
\beq
\chi_A:X\rightarrow \{1,0\},
\label{eq:chiA}
\eeq
defined by
\beq
\chi_A(x)=\left\{\begin{array}{ll}
		1	& \mbox{if $x\in A$}\\
		0	&\mbox{otherwise.}
\end{array}
\right. 
\label{eq:chi}
\eeq
or, diagramatically,
\[
  \begin{array}{c}\mbox{\epsfig{file=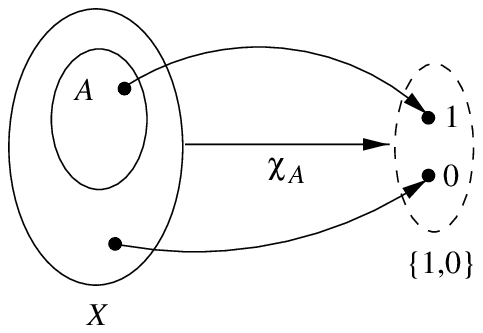}}\end{array}
\]
The function $\chi_A$ splits the elements of $X$ into those that also 
belong to $A$ (which get the label 1) and those that do not
(which are marked 0). The subset $A$ can be recovered using
the inverse of the characteristic function,
as the inverse image
\beq
A=\chi_A^{-1}(1).\label{eq:chi-1}
\eeq

What we have in mind for our time-dependent setup is 
a generalised notion of a subset, appropriate for evolving 
sets.  In order to 
generalise, let us first scrutinize the above construction. 
Let us also treat sets as objects in the category ${\bf Set}$,
since we will later have to use evolving sets as objects in the 
apropriate category. 

As we have already said above, the category ${\bf Set}$ has
sets as its objects and its arrows are set functions. 
Of course, sets contain {\it elements}, and this is essential 
in the above definition of a subset. On the other hand, 
when we lay out the category ${\bf Set}$ there is nothing 
about elements to start with. How do they come about?

{\bf Set} is a category with a {\it
terminal object}. A terminal object in a category
has the property that there is a unique arrow from any other object 
in the category to it. If it exists, the terminal object is unique up to 
isomorphism.\footnote{
In {\bf Set} there is also an {\it initial object},  a set that is special 
in that there is a unique arrow from it to every other set in the category.
With some thought we can convince ourselves that this is the empty 
set.}
In {\bf Set}, the terminal object should then be a set that is special in
that there is a unique function from every other set into it. We can
check that this is the one-element set in the following way. 
The function $g(x):=\sl{constant}$ on an element 
$x\in X$, exists for any set $X$. Since its only output is $\sl{constant}$,
for each $X$ the function $g$ is unique (up to isomorphism).
Thus, the terminal object is the set with the single 
element $\sl{constant}$. We will denote the terminal object  $\{0\}$.
 
Note that we only care about the fact that the terminal object
has only one element and not about 
which one this element is since all one-element sets are isomorphic
and our constructions only need to be good up to isomorphism.
We therefore denote it by $\{0\}$ 
but we could just as well 
have used $\{1\}$, $\{e\}$ or $\{\sl{constant}\}$
(and the 0 in the terminal object has absolutely nothing
to do with the 0 in the set $\{1,0\}$).

Even if we are
not told anything about the objects of ${\bf Set}$ having elements, 
we can infer that they do
from the existence of a terminal object.
Consider a function from $\{0\}$ to some set $X$. Since the range
of this function is a single element, the only output it can have is 
an element of $X$. Thus, the functions from $\{0\}$ to $X$
are in one-to-one 
correspondence with the elements of $X$, and we can, in fact, define
an element of $X$ to be such a function. 

This may seem a far too complicated way of doing things, and it certainly
is if all we cared about were sets. We need the concept of elements since we 
need to tell which events in $\C$ have occured at any given time. 
The above definition of an element proves useful in the rest of 
the present work because it can be generalised to evolving sets.  
  
Going back to the subset $A$, we have 
\beq
A\stackrel{\subseteq}{\longrightarrow}X\stackrel{\chi_A}
{\longrightarrow}\{1,0\}.
\eeq
The first thing to note is that when we refer to $A$ as 
the subset of $X$ we mean that there is an inclusion function from $A$
to $X$. To emphasize this, we will give the name $f$  
to this function,
\beq
A\stackrel{f}{\longrightarrow}X\stackrel{\chi_A}
{\longrightarrow}\{1,0\}.
\label{eq:ssd}
\eeq
The characteristic function  $\chi_A$ is an 
equivalent description of the function $f$. 

We choose to interpret 1 as 
``it is true that $x$ is in $A$'', i.e.\ 1 is chosen
as the {\it true} 
one of the two truth-values of the set $\{1,0\}$.\footnote{
We are effectively replacing the set $\{1,0\}$ by the
set $\{\mbox{\sl true}, \mbox{\sl false}\}$ of boolean truth values.}
``Choosing the element  1'' from  $\{1,0\}$ as described above,
means that we use a
function from the terminal object to the set $\{1,0\}$ that outputs 
the element 1,
\beq
T:\{0\}\longrightarrow\{1,0\}\qquad\mbox{such that}
\qquad T(0)=1.
\label{eq:T}
\eeq
Its name $T$ is suggestive of ``true'', since this function represents our 
choice for which element of $\{1,0\}$ makes the expression 
``$x\in A$'' be true.

In its full detail then, the set inclusion diagram  (\ref{eq:ssd}) is
therefore
\[
  \begin{array}{c}\mbox{\epsfig{file=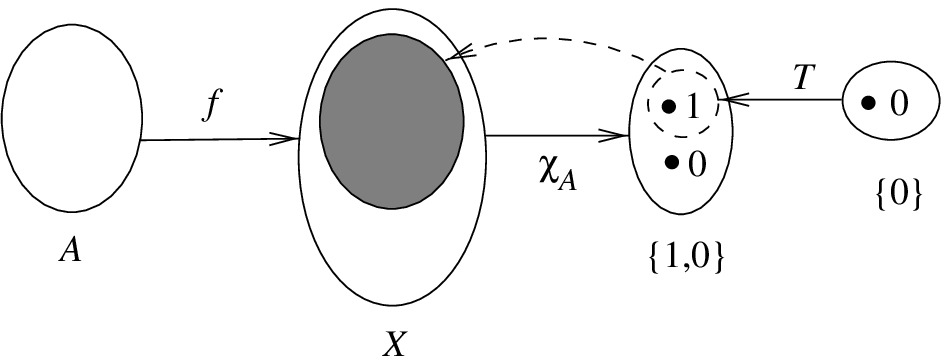}}\end{array}
\label{AfXdiagram}
\]
The shaded set is the image of $f$ of $A$ in $X$, 
or, equivalently, the inverse image $\chi^{-1}_A(1)$
in $X$.

\subsection*{A subset is a pullback}
\label{pullback}

What the characteristic function $\chi_A$ does is pick the elements of
$X$ that carry the label 1 from the two possible in $\{1,0\}$. The subset $A$ 
is then defined to be the set that contains those elements only. 

The philosophy here can be roughly interpreted as follows. The set of truth-values $\{1,0\}$ is the simplest set we have that, together with the simplest possible 
(but non-trivial) inclusion function $T:\{0\}\rightarrow \{1,0\}$, 
exhibits all the features of a subset of a set.
The characteristic function $\chi_A$ ``lifts off'' this model case of a subset, 
applies it to the particular set $X$ we provide, and returns the subset $A$.

The mathematical explanation of the same philosophy is a 
{\it pullback diagram}.
We group the functions $f:A\rightarrow X$ of eq.\ (\ref{eq:ssd}), 
$\chi_A:X\rightarrow\{1,0\}$ of eq.\ (\ref{eq:chiA})
and $T:\{0\}\rightarrow \{1,0\}$ 
of equation (\ref{eq:T}) into the diagram
\beq
\bundle{A}{f}{X}
\bundlemap{}{\chi_A}
\bundle{\{0\}}{T}{\{1,0\}}
\eeq
where the top arrow is the unique function from $A$ to $\{0\}$ (the function 
$g$ we defined above). 

We know that $A$ contains those $x\in X$ with $\chi_A(x)=1$. This is
precisely the statement that the diagram above is a pullback square, 
i.e. that $A$ is the set that contains $x\in X$ that have the same image 
under $\chi_A$ as $T(0)$,
\beq
\chi_A(x)=1=T(0).
\eeq
We say that $A$ is the {\it pullback}
in the above diagram. In terms of functions, 
$\chi_A$ is the only function from $X$ to $\{1,0\}$ along which 
the $\it{true}$ function
$T$ pulls back to yield the inclusion function $f$, i.e.\
$f$ is the pullback of 
$T$ along $\chi_A$.\footnote{A subset is a special case of a pullback. 
In general, that $A$ is 
a pullback means it  contains {\it pairs} of elements $(x,0)$ with 
$x\in X$ and  $0\in\{0\}$ that have the same image in $\{1,0\}$. 
It is because $A$ is a subset that 
the second argument in this pair is the single element 0 of the 
terminal object and, since we can unambiguously drop it, we think of $A$ as
containing elements of $X$ only.}

At this stage we are done with all the technical material from set theory.
We will now go ahead to apply it, generalise it, and draw conclusions
from it. The generalisation will attempt to capture the following:
In the case of sets and subsets that we have seen, it is 
characteristic that $\chi_{A}$ splits the original set in exactly two 
parts, those that belong to the subset and those that do not, 
with anything in the middle excluded. In such cases
we only need two truth-values and $\{1,0\}$ suffices.
For example, this defines the causal past of a {\it single} 
event in the causal set: we can assign 1 to all events $r\leq p$ 
and 0 to all $q\geq p$. 

However, we aim for more. We wish to have truth-values that inform us,
not only whether an event has occured or not, but also how 
long we need to wait till it does, or, whether it will not happen no matter
how long we wait. This is provided by the enlarged set of truth-values
(in fact a functor) and the characteristic function  for the evolving set $Past$.

\section{The evolving set $\Past$}
\label{past}
The ``viewpoint'' of an event $p$ in the causal set, the history 
of the world to the knowledge of $p$, is the
set of events in the causal set in the past of $p$. This
is a subset of the whole causal set $\C$. 
Here we consider the causal pasts 
{\it evolving over the causal set}.

As we said in section \ref{outline}, 
pasts of events in $\C$ are  the outputs of a functor
from the causal set to sets,
\beq
\Past:\C\longrightarrow{\bf Set}
\eeq
that assigns to each $p\in\C$ its past
\beq
\Past(p)=\left\{r\in\C:r\leq p\right\}.
\eeq
and to each causal relation $p\rightarrow q$, when it appears in 
the causal set, a function
\beq
\Past_{pq}:\Past(p)\longrightarrow \Past(q),
\eeq
which includes the idenity map $\Past_{pp}:\Past(p)
\rightarrow\Past(p)$.

As an example, here is a causal set 
\beq
 \begin{array}{c}\mbox{\epsfig{file=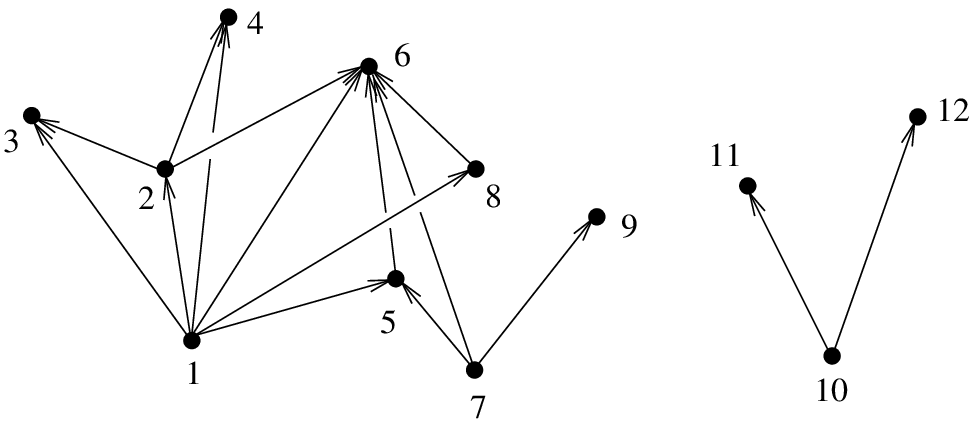}}\end{array}\nonumber
\label{eq:smallcausalset}
\eeq
where we have drawn all the causal relations (and not only the 
nearest-event ones as is often done).
When $\Past$ varies over this causal set, it gives
the sets of causal pasts
\beq
 \begin{array}{c}\mbox{\epsfig{file=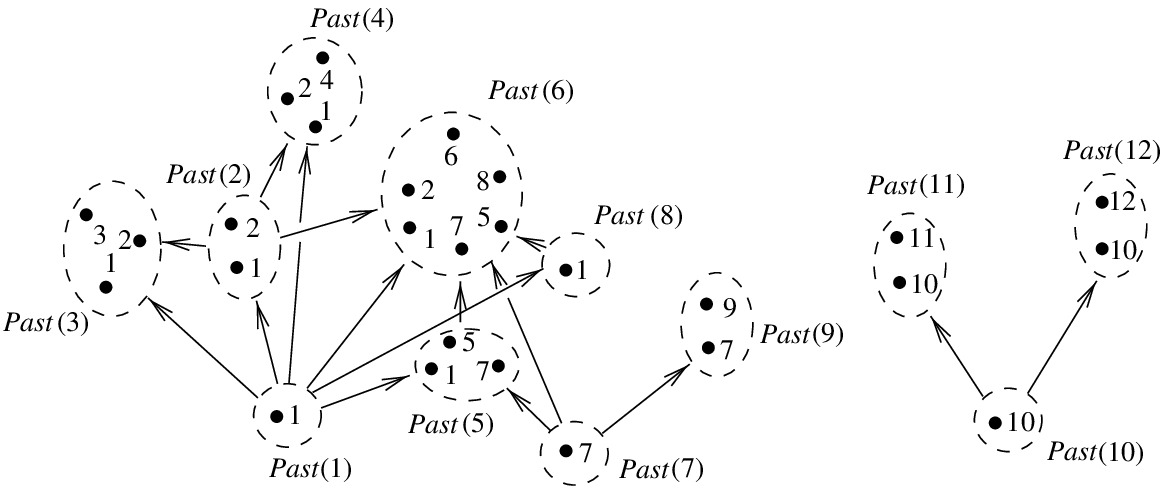}}\end{array}\nonumber
\eeq
which themselves form a partially ordered set as
the arrows are set inclusions. 

$\Past$ is a $\C\rightarrow{\bf Set}$ functor. 
It is a particular object in the category of functors
from a partially ordered set $\C$ to the category of sets
${\bf Set}$,  denoted by 
${\bf Set}^\C$. (Arrows between these functors are natural 
transformations, an example of which we define below.)

Such functors, since they are evolving sets,
are generalisations of sets. For this reason, the category 
${\bf Set}^\C$ has common features with the category ${\bf Set}$ 
of sets. It, also, has a terminal object, and it is a 
generalisation---in fact, an evolving version---of 
the one-element set that is the terminal object for sets. 
It is the functor $t.o.$ which assigns to each 
$p\in\C$ the one-element
set $\{0\}$
\beq
t.o.:\C\longrightarrow {\bf Set}\qquad\mbox{with}\qquad
t.o.(p)=\{0\}.
\label{eq:to}
\eeq
with an identity arrow between one-element sets for 
every existing causal link,
\beq
t.o._{pq}:\{0\}\longrightarrow\{0\}\qquad
\mbox{when}\qquad p\leq q.
\eeq
We can visualise the terminal object as the same web of relations as in 
$\C$, but with $\{0\}$ stuck in place of every event:
\beq
 \begin{array}{c}\mbox{\epsfig{file=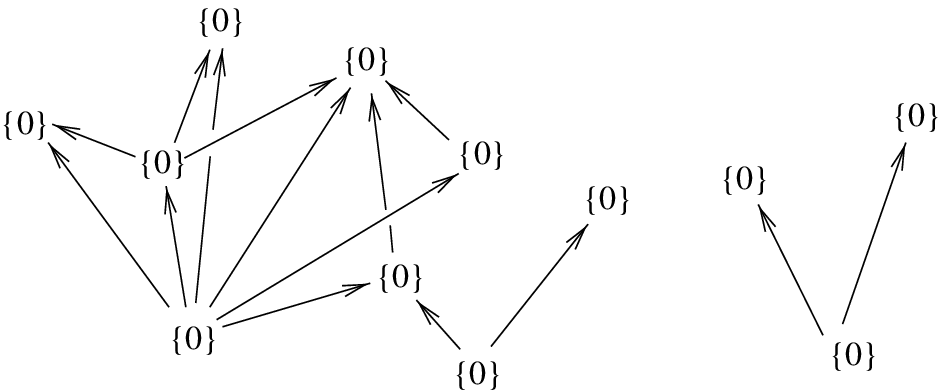}}\end{array}\nonumber
\eeq

What the causal set encodes is whether one event is before or
after some other event. Let us examine this.      
Suppose we are at event $p$ and need to say whether some 
other event $r\in\C$ has occured. Clearly, if $r\leq p$, then
$r$ has happened. We can, however, do better than this and
assign truth-values at $p$, not only to ``$r$ precedes 
$p$'', but to {\it when}\/ $r$ {\it will happen}.  Suppose 
there is some $q$ in the future of $p$, $q\geq p$,
for which $r$ is past, $r\leq q$.  At $p$, the ``time'' we need to
wait before $r$ occurs is the causal relation 
$p\rightarrow q$.\footnote{An example of ``length of time''
truth-values is Newtonian 
histories, which we work out in the next section. There, the 
arrow $p\rightarrow q$ can be assigned a length, the number of 
events between $p$ and $q$. This, of course, is only possible
because in the Newtonian case the causal set is a full linear 
order. In general, it is the arrow itself that 
we think of as the ``time elapsed'' between $p$ and $q$.}
This arrow is therefore a candidate for the truth-value, at $p$,
for $r$ to have occured. 

Under more careful inspection, we can see that a single arrow is 
insufficient. First, if $q$ has both $p$ and $r$ in its past, 
so will an event $q'\geq q$. That is, the causal relations 
$p\rightarrow q'$ for all $q'\geq q$ should also be included
in the truth-value we are calculating. Second, in general, there 
will not be a single first event $q$ in the common future 
of $p$ and $r$, but a set of such events, as is the case in the diagram 
below.
These, and all events in their future, should be the causal 
relations in the truth-value at $p$ for $r$ having occured. 
They are the bold arrows in the following diagram.

\beq
 \begin{array}{c}\mbox{\epsfig{file=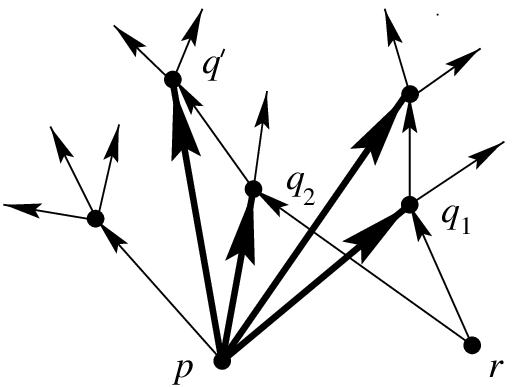}}\end{array}\nonumber
\label{eq:sieve}
\eeq

Let us formalize this. Let us define
$R(p)$ as the set of 
all causal relationships that start at $p$ (including $p\leq p$),
\beq
R(p)=\{\mbox{all }p\longrightarrow p':p\leq p'\mbox{ in }\C\}.
\eeq

Next, we define subsets $S(p)$ of $R(p)$ 
such that, if $S(p)$ contains the causal relation $p\rightarrow q$, 
it should also contain $p\rightarrow q'$, for every $q'\geq q$.
$S(p)$ is therefore a left-extendible subset of $R(p)$
since the above definition can also be read as,
if $(p\rightarrow q)\in S(p)$, and there is $(q\rightarrow q')\in \C$,
then $(p\rightarrow q')=(q\rightarrow q')\circ(p\rightarrow q)\in S(p)$.
Namely, $S(p)$ is closed under left multiplication (extending a causal 
relation to the future).

A left-extendible subset $S(p)$  of $R(p)$ (in our convention\footnote{
We call a sieve what in \cite{McLM} is a cosieve.  })
is called a {\it sieve on} $p$. The truth-value at $p$ that 
we are looking for is precisely a sieve. 
In fact, since it is not simply a yes/no truth-value, but also 
tells us at $p$ when $r$ will occur, it   is called a {\it time-till-truth
value}. A sieve is the type of truth-value
appropriate to statements which, {\it once they become true, they remain
true}, as is the case for an event having occured. 

Therefore, at $p$, we have a sieve truth-value for every other event $r$ 
in the causal set. The set of all these sieves is 
the set of truth-values at $p$. We call this set $\Omega(p)$, 
\beq
\Omega(p)=\{\mbox{sieves on }p\}.
\label{eq:Omegap}
\eeq

A special subset of the events in $\C$ is, of course, those in $\Past(p)$. 
There is a sieve in $\Omega(p)$ that corresponds to these 
particular events. They are special because at $p$ they have already 
happened. While we may think of the sieve truth-value in $\Omega(p)$ 
for $r\notin\Past(p)$ as a {\it partially true} value, the sieve
for the events in $\Past(p)$ is the {\it totally true} value. 
We can use this understanding to find the sieve in $\Omega(p)$ 
that corresponds to
the events in $\Past(p)$ and thus the set $\Past(p)$ itself. 
Since these events have occured at $p$ and at all events that follow
$p$, the truth-value we are looking for is the entire $R(p)$, the 
{\it maximal sieve} in $\Omega(p)$.  

In more detail, $R(p)$ is an element of $\Omega(p)$ which is selected
in the manner described in section \ref{sets} for choosing
elements of a set using the terminal object.
We define an inclusion of  $\{0\}$ into $\Omega(p)$,
which we call the {\it totally true} arrow at $p$,  $T_p$:
\beq
T_p:\{0\}\longrightarrow \Omega(p),
\eeq
such that
\beq
T_p(0)=R(p)=(\mbox{maximal sieve on }p).
\eeq
Let us now observe that  $\Past(p)\subset\C$ and call this inclusion $f_p$,
\beq
f_p: \Past (p)\subseteq\C .
\eeq
Both $\Past(p)$ and $\C$ are sets, and as we saw in section \ref{pullback},
that the former is a subset of the latter means that 
$\Past(p)$ is the pullback in the diagram
\beq
\bundle{\Past(p)}{f_p}{\C}
\bundlemap{}{\chi_p}
\bundle{\{0\}}{T_p}{\Omega(p)}
\eeq
namely,  $Past(p)$ contains those events $r\in\C$ that satisfy 
\beq
\chi_p(r)=(\mbox{maximal sieve on }p)=T_p(0),
\eeq
exactly as we saw in section \ref{sets}. 

We have, therefore, found that the set of truth-values $\Omega(p)$ at a 
given event $p$ is the set of sieves on $p$ and, 
in particular, that the maximal sieve gives 
the component $\Past(p)$ of $\Past$ at $p$. We next have to let 
these results ``vary'' over $\C$ so that we can define the functor 
$\Past$ by a generalisation of what we did for $\Past(p)$. 

As we move from $Past(p)$ to $Past(q)$,
the appropriate truth-values are consistently given by $\Omega(p)$ and 
$\Omega(q)$, so that we can find $Past(q)$ given $Past(p)$,  
$p\leq q$ and $\C$.
That is, the diagram
\beq
\bundle{\Past(p)}{{\Past_{pq}}}{\Past(q)}
\bundlemap{f_p}{f_q}
\bundle{\C}{\mbox{id}_\C}{\C}
\label{eq:Cdiagram}
\eeq
commutes (id$_\C$ is the identity operation on the causal set).
On the left of the diagram we have components of the functor $\Past$. 
We can organize the righthand side in the same way by
constructing the {\it constant} functor
\beq
\World:\C\longrightarrow {\bf Set}
\eeq
with 
\bea
\World(p)&=&\C\qquad\mbox{for any }p\in\C,\\
\World_{pq}&=&\mbox{id}_\C.
\eea
Now the diagram (\ref{eq:Cdiagram}) reads 
\beq
\bundle{\Past(p)}{\Past_{pq}}{\Past(q)}
\bundlemap{f_p}{f_q}
\bundle{\World(p)}{\World_{pq}}{\World(q)}
\label{eq:pastpullback}
\eeq
What this diagram says is that $\Past$  is a {\it subfunctor} of $\World$,
i.e.\ there is an inclusion 
\beq
f:\Past\longrightarrow \World, 
\eeq
which is  a natural transformation, i.e. it has components 
$f_p, f_q$ that make the diagram (\ref{eq:pastpullback})
commute, $f_q\cdot\Past_{pq}=\World_{pq}\cdot f_p$ 
for every two events  $p,q$ that are causally related.

The set of truth-values for $Past(p)$ is $\Omega(p)$ while for $\Past(q)$
it is $\Omega(q)$. These, too, are components of the functor
\beq
\Omega:\C\longrightarrow {\bf Set}
\eeq
that for each event in $\C$ outputs its set of truth-values, the sieves on that
event, and has functions $\Omega_{pq}:\Omega(p)\rightarrow \Omega(q)$
that, given the set of sieves on $p$ give the set of sieves at $q$. 

We, therefore, have a characteristic function $\chi$ from $\World$ to 
$\Omega$,
\beq
\chi:\World\longrightarrow \Omega.
\label{eq:chiPast}
\eeq
which is also a natural transformation, it has components at each event that 
compose as $f$ did above.
We obtain the particular subfunctor $\Past$, when, at each event in 
the causal set, $\chi$ maps into the maximal sieve at that event. 
Namely, the function $T_p$ above is the $p$-component of 
the arrow $T$ from the terminal object $t.o.$ for ${\bf Set}^\C$ 
defined in (\ref{eq:to}) to $\Omega$,
\bea
T&:&t.o.\longrightarrow\Omega,\qquad\mbox{with}\\
T_p&:&t.o.(p)=\{0\}\rightarrow \Omega(p).
\eea

We, finally, have everything we need to give the subfunctor $Past$ as 
a pullback.
All four $\Past$, $\World$, $t.o.$ and $\Omega$ are objects in the category 
${\bf Set}^\C$. We may then construct the diagram
\beq
\bundle{\Past}{f}{\World}
\bundlemap{}{\chi}
\bundle{t.o.}{T}{\Omega}
\eeq
and ask that the functor $\Past$ makes this diagram pullback.
This means that it reduces to the 
diagram (\ref{eq:Cdiagram}) consistently at each event in the causal set.
$\Past$ is a subfunctor of $\World$.\footnote{
All of $\Past$, $t.o.$, $\World$ and $\Omega$ are evolving sets, and 
$\chi$, $T$ and $f$ are natural transformations.  In particular, 
$\chi$ has components 
\beq
\bundle{\C}{\mbox{id}_\C}{\C}
\bundlemap{\chi_{p}}{\chi_{q}}
\bundle{\Omega(p)}{\Omega_{pq}}{\Omega(q)}
\eeq}

We now know how to specify an evolving set.
We have defined an evolving causal past by giving the functor $\Past$ that 
generates the causal past of each event
and corresponding evolving set of truth-values $\Omega$. 
We can now  analyze their physical significance. 

Overall, we have achieved two things. One is to express the 
causal structure in terms of causal pasts at each event, which 
we regard as physically more satisfactory than an 
entire spacetime or causal set. This is the description that can be given from
an observer inside the causal set. 

Second, even when the causal set is not a lattice, the evolving sets 
over it is one.  As we will see next, they satisfy
a particular algebra, called a Heyting algebra, whose
operations reflect the underlying causal set. Thus, 
we can give the causal structure {\it algebraically}. 
We do this in the section \ref{algebra}.  

Before we continue with the algebra of causal sets,  we will 
give a simple example of causal pasts in a causal set. We will 
consider histories in a discrete Newtonian universe.

\section{Discrete Newtonian histories}
\label{newtonian} 

Discrete Newtonian time evolution  
is the case where integers 
$0,1,2,3,\ldots\in{\bf N}$ can label the preferred time parameter.  
In a Newtonian world, we have a preferred foliation 
with slices $S_0, S_1, S_2, S_3, \ldots$ labelled by the 
time stage when they occur. It is, therefore, a very special causal 
set: the fully ordered set of integers.

At some given time $n$, we may ask for the history up to that time.
It is the sequence of slices up to $n$,
namely,\footnote{
In Newtonian evolution the speed of light is infinite and we do not have 
the local (bubble)  evolution present in a causal universe. Effectively, 
the slice
$S_n$ can simply be replaced by $n$, or, equivalently, we can 
reduce the slice to contain a single event and the causal structure
will not change. Therefore, the discrete Newtonian history 
$\{S_0, S_1, S_2, \ldots, S_n\}$
is isomorphic to just  $\{0,1,2,\ldots\,n\}$.}
 
\beq
\History(n)=\left\{S_0, S_1,S_2, \ldots, S_n\right\}.
\label{eq:hist_n}
\eeq
Clearly there is a history at each 
time stage  $n\in{\bf N}$. We can set up a functor $\History$,
\beq
\History:{\bf N}\longrightarrow{\bf Set},
\eeq
 which, when fed the time
instant, spits out the history up to that time,
 $\History(n)$,  as
given by (\ref{eq:hist_n}). The functor $\History$ also 
provides maps between histories which are just the inclusions
of the shorter  histories into the longer ones,
\beq
\History_{nm}:\History(n)\subseteq \History(m)
\qquad\mbox{whenever}\ n\leq m.
\eeq
 
We may now want to ask whether some slice $S_i$ is already in 
$History(n)$ and, if not, when it will be included. 
First, we note that all slices can be found in the components of the
constant functor 
\beq
\World:{\bf N}\longrightarrow{\bf Set},
\eeq
since its components are
\beq
\World(n)=\left\{S_0,S_1,S_2,\ldots\right\}
			\quad\mbox{for any }n\in{\bf N}.
\eeq
$History$ is a subfunctor of $World$ since, at each time $n$,
$History(n)\subseteq World(n)$,
\beq
\bundle{\History}{f}{\World}
\quad\mbox{:}\quad
\bundle{\History(0)}{\subseteq}{\World(0)}
\bundlemap{}{}
\bundle{\History(1)}{\subseteq}{\World(1)}
\bundlemap{}{}
\bundle{\History(2)}{\subseteq}{\World(2)}
\bundledots{}{}
\eeq
(By analogy to the section on sets, $\World$ is the 
evolving set version of the set $X$, while $\History$ 
is an ``evolving subset'' in place of $A$.)

Diagramatically, we have
\beq
 \begin{array}{c}\mbox{\epsfig{file=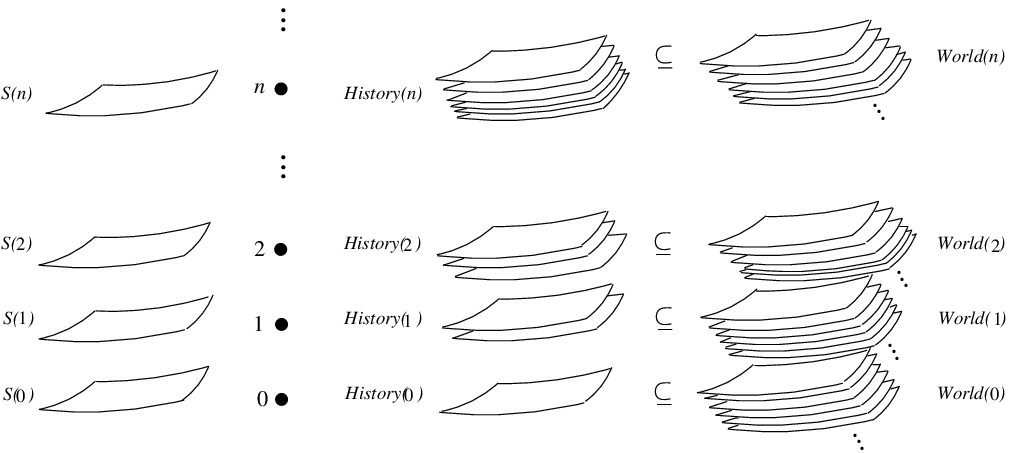}}\end{array}\nonumber
\eeq

At time $n$, the slice $S_i$
may or may not be in $\History(n)$. If it is not  already history
at $n$, then, if we wait for some time $t$ we may find that 
$S_i\in \History(n+t)$. As a result, the characteristic function
that tells us at $n$ when $S_i$ is history is
\beq
\chi_n(S_i)=\left\{\begin{array}{ll}
		\mbox{least }t &\mbox{at which } S_i\in \History(n+t),
			\mbox{if this occurs}\\
		\infty	&\mbox{otherwise.}
\end{array}
\label{eq:chiN}
\right. 
\eeq
Of the numbers that $\chi_n(S_i)$ spits out, 0 is to be 
understood as ``true now'' ($S_i$ is already history at time $n$),
1 is ``true tomorrow'' and so on. $\infty$ is the symbol we chose for 
``never true'', if the particular slice $S_i$ never appears
in $\History$.

Note that $\chi_n$ takes its values in the set ${\bf N}_\infty$,
the integers together with $\infty$. Therefore, 
the {\it set of truth-values at $n$} is 
${\bf N}_\infty$. In fact, the set of truth-values
is ${\bf N}_\infty$ at any time. We can chain these sets
together to form $\Omega:{\bf N}\longrightarrow {\bf Set}$,
that represents the time-development of the set of truth-values,
only, in our especially simple case, there is no time development
and
\beq
\Omega(n)={\bf N}_\infty\quad
	\mbox{for any }n\in{\bf N}. 
\eeq
$\Omega$ then is the ``constant evolving set'' (with identity maps 
between components)
\beq
\Omega=\left\{{\bf N}_\infty,
		{\bf N}_\infty,
		{\bf N}_\infty,
	 {\bf N}_\infty,\ldots\right\}.
\eeq

Note that the {\it true now} (at $n$) output of $\chi_n$ in
(\ref{eq:chiN}) is the value 0. This is the maximal sieve in 
$\Omega(n)$ above. We can see that this is because the sieves here
tell us how long we need to wait before $S_i$ joins $\History$,
and 0 is, of course, the shortest possible wait. It is the maximal 
sieve since it contains all longer waits, $1,2,\ldots,n,$ etc.
\footnote{
We can write $\Omega(n)$ as all the arrows from $n$ to 
any $m\geq n$, i.e.\ all the possible ``waiting times'', or 
time-till-truth values:
\beq
\Omega(n)=\left\{\begin{array}{lllllll}
	n\longrightarrow n\\
	n\longrightarrow n+1\\
	\vdots\\
	n\longrightarrow n+k\\
	n\longrightarrow n+k+1\\
	\vdots\\
	n\longrightarrow\infty
\end{array}\right. .
\label{eq:Nsieves}
\eeq
The length of these arrows is the number that $\chi_n$ of 
eq.\ (\ref{eq:chiN}) outputs. 
These are the sieves on $n$. Going from $\Omega(n)$ to $\Omega(n+1)$,
the length of each arrow goes down by one since, what at $n$ was
``true tomorrow'', at $n+1$ is ``true now'' and so on.
Thus $\Omega$ is closed under making the arrow $n\rightarrow m$ 
even longer, which means that what is true now will remain true tomorrow.
The arrow $n\rightarrow n$ is the shortest one, it means no wait, 
i.e.\ true now (at n).}

Finally, the history at a given time can be obtained
by first defining the function $T_n$ at $n$ to have output 0
which is the {\it true now} value in (\ref{eq:chiN}).
\beq
T_n:\{0\}\rightarrow\Omega(n)\qquad\mbox{with}\qquad
		T_n(0)=0.
\eeq
Then $History(n)$ is the pullback in
\beq
\bundle{\History(n)}{\subseteq}{\World(n)}
\bundlemap{}{\chi_n}
\bundle{\{0\}}{T_n}{\Omega(n)}
\eeq
namely, it contains those slices $S_i$ in $\World(n)$ that satisfy
\beq
\chi_n(S_i)=0=T_n(0),
\eeq
and thus are already true at $n$.

\section{The complement of $\Past$ measures the causal 
structure of $\C$}
\label{complement} 

In section \ref{past} we provided a time-dependent way of 
telling when some event in a causal set has occured (equations 
(\ref{eq:chiPast}) and (\ref{eq:Omegap})). 
What characterises an event that will {\it never} happen?
In this section we give the relevant definition which 
encodes much of the causal structure of $\C$. We then make some 
first remarks about the possibility of defining algebraically 
global properties of a causal set, such as black holes and branchings.

A sieve on $p\in\C$ is the time-till-truth value for some other
event $r\in\C$ to happen. In the same way, if $r$ will never 
happen, then it should satisfy
\beq
r\notin\Past(q)\qquad\mbox{for any }q\geq p,
\label{eq:r}
\eeq
since $r$ never happening for $p$ means that 
it will never be in the past of the future of $p$.

The set of events satisfying (\ref{eq:r}) is the {\it evolving 
set complement} of $\Past(p)$, the set of events in the causal set
$\C$ that are not in the past of the future of $p$:
\beq
\neg\Past(p)=\left\{r\in\C:r\notin\Past(q)\mbox{ for any }q\geq p
	\right\}.
\label{eq:negpast}
\eeq 

As a first example, consider the Newtonian case. There the 
functor $\History$ plays the same role as $\Past$, and thus
\beq
\neg \History(n)=\{k\in {\bf N}: S_k\notin \History(m)\mbox{ for any }
				m\geq n\}.
\eeq
But, since the causal ordering 
is a full linear ordering, any slice will at some point 
join $\History$.
Thus, the complement of $\History(n)$ is {\it empty}. 

It is generally the case with a causal set that is a lattice that 
the complement of $\Past$ is empty. This is because, by definition,
when $\C$ is a lattice any two events $p$ and $r$ have a 
least upper bound which is an event with both $p$ and $r$ in its past.
Thus, definition  (\ref{eq:negpast}) returns the empty set.
Note that this implies that $\neg\Past$ cannot distinguish between
different lattice causal sets.

Second, let us take the very small causal set in (\ref{eq:smallcausalset}).
It is not a lattice.
There is no single ``big bang'' event and no single ``final crunch''.  
Consider the event labelled 7 and let us find the complement of 
$\Past(7)$.
$\neg \Past(7)$ is all the events not in the past of the 
future of 7, i.e.\ 3, 4, 10, 11 and 12. We mark them white:
\beq
 \begin{array}{c}\mbox{\epsfig{file=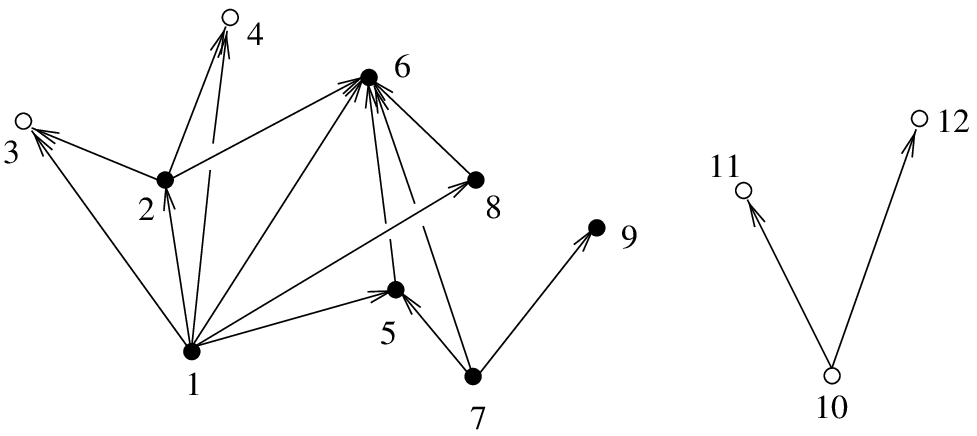}}\end{array}\nonumber
\eeq

Let us imagine placing an observer on event 7. The future of 7
is 5,6 and 9. In these future times, he will receive information 
from anything in the past of 5,6 and 9, i.e.\ from 1, 2, 5 and 8.
He will never, however, be able to see 3, 4, 10, 11 and 12,
the events in $\neg \Past(7)$. As far as 7 is concerned, 
$\neg \Past(7)=\{3,4,10,11,12\}$ is {\it beyond its causal horizon},
7 will never receive information from these events. 

It is not suprising that there is such a boundary when the causal 
structure is nontrivial, and parts of the causal set are either 
disconnected or branch off. This is reminiscent of  
topology change in a spacetime, although, naturally, we do not 
have the same notion of topology here. 

We should emphasize that this is a horizon for 7 and not a
global property of $\C$. 
Such an event-dependent notion of causal horizon is useful in 
constraining a microscopic theory to reproduce the Einstein 
equations at large scale (used in \cite{TJ}). However, since this paper
is devoted simply to the exposition of the
evolving set $\Past$ 
and not its applications, we will not go further
into this, except for comments in the conclusion section. 

A global property of the spacetime, such as a branching of the 
causal set or a black hole, needs, 
presumably, the agreement of the complements of a
large class of events. 
A permanent braching of $\C$ will result to all events in one branch
having (almost all) the events in the other branch in their complements.
It is a very interesting but subtle problem to express a black hole in 
terms of the causal structure by considering intersections
of the complements of the events in $\C$. This is work currently 
in progress. 

The non-standard complement of $\Past$ is one of the four 
operations of the algebra of (evolving subsets of) evolving sets. 
It is a {\it Heyting algebra}, which we define in the next section 
by giving its four opreations.

\section{Evolving sets obey a Heyting algebra}
\label{algebra}    
 
Recall  that $\Past$, as a functor from $\C$ to sets, can be regarded
as an evolving set, a set that varies over all events in $\C$. 
As with the other definitions in this paper, we start from sets
and generalise to evolving sets. 

\subsection{The algebra of sets}
\label{setalgebra}

Take a set $X$, and consider the set of its subsets, its {\it powerset}.
There are four possible operations on the subsets of $X$, 
and they produce four unique new sets:
\begin{itemize}
\item
Union: $(A\cup B)=\{x\in X: x\in A$ or $ x\in B\}$.
\item
Intersection: $(A\cap B)=\{x\in X: x\in A$ and $x\in B\}$.
\item
Implication: $(A\Rightarrow B)=\{x\in X:$ if $x\in A$ then $x\in
B\}=\{x\in X: x\notin A$ or $x\in B\}$.
\item
Complement: \\
The complement $\neg A$ of $A$ is characterised by 
\beq
A\cap\neg A=\emptyset\qquad\mbox{and}		
		\qquad A\cup\neg A=X,
\label{eq:setcomplement}
\eeq
and, therefore, $\neg A=\{x\in X: x\notin A\}$.
\end{itemize}

Note that the complement can, in fact, 
be derived from the implication operation by defining
\beq
\neg A=(A\Rightarrow \emptyset).
\label{eq:setcompfromimpl}
\eeq
This means that $\neg A$ contains those $x\in X$ which
``if they are in $A$ they also are in $\emptyset$'', 
which is a formal way to exclude the elements of $A$ from 
$\neg A$. As on other occasions in this paper, it is this
twisted definition that we generalise below. 
From the two relations (\ref{eq:setcomplement}), one can 
prove the familiar
\beq
\neg\neg A=A.
\label{eq:setnegneg}
\eeq
 
Diagramatically, the four set operations are:
\beq
 \begin{array}{c}\mbox{\epsfig{file=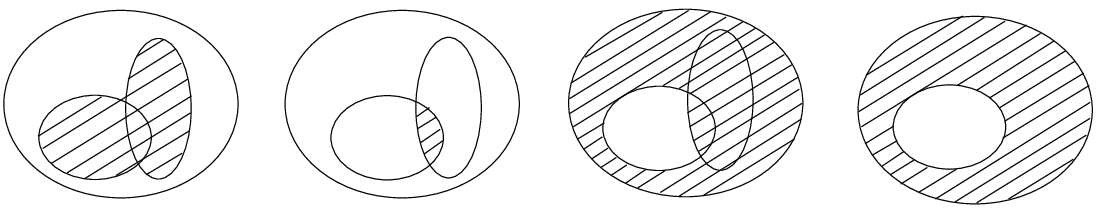}}\end{array}\nonumber
\eeq

The partially ordered set of subsets of $X$ is a 
lattice, since every two subsets have their
union as a l.u.b.\ and their intersection as a g.l.b.
The infimum of the powerset of $X$ is the empty set $\emptyset$ 
and the supremum is $X$ itself. Importantly, it is a 
distributive lattice, i.e.\ the identity
\beq
A\cap(B\cup C)=(A\cap B)\cup(A\cap C)
\eeq
holds. 

A distributive lattice in which every element has a complement
that satisfies (\ref{eq:setcomplement}) is a Boolean lattice, or,
a {\it Boolean algebra}. Therefore, the subsets of $X$ form a 
Boolean algebra.

\subsection{The algebra of evolving sets}
\label{evsetalgebra}

With evolving sets, we have the same four operations of 
union, intersection, implication and complement, but
different relations between them and thus a different algebra
which is known as a Heyting algebra. 

Consider two evolving sets, or functors from $\C$ to ${\bf Set}$
which  are subfunctors of our constant functor $\World$.
Let us call them $F$ and $G$. We have worked with one of them,
$\Past$. Another example would be the functor generating events in the 
future of each event in $\C$, or sets of spacelike separated events 
in each causal past, etc.

The four possible operations on $F$ and $G$ are given componentwise 
(at each event in $\C$) but so that they preserve the causal 
structure of $\C$. They are:\footnote{
It is somewhat difficult to draw diagrams for evolving sets. 
If the reader wishes to have an illustration of a Heyting 
algebra, 
a good alternative is to draw open sets. Open subsets of a given 
(open) set $O$ also satisfy a Heyting algebra.  
Given an open set $U\subset O$, the set-theoretic complement 
of $U$  (i.e.\ the one satisfying the definition 
(\ref{eq:setcomplement})), which we will call $U^c$ to avoid confusion, 
is a closed set. If we want an algebra of open sets, we need,
instead of $U^c$, to use the interior of the set-theoretic complement of 
$U$, $Int(U^c)$. But then, clearly, $U\cup Int(U^c)\subset O$ 
since the closure of $U$ has been left out.
Open sets are the standard example of a Heyting algebra in 
the literature.
That they behave in the same way as evolving sets is not surprising,
roughly speaking, they 
both involve an infinite sequence, the former of time stages to the
future, the latter
of points to the boundary.}

\begin{itemize}
\item
Union: $(F\cup G)(p)=F(p)\cup G(p)$.
\item
Intersection: $(F\cap G )(p)=F(p)\cap G(p)$.
\item
Implication: 
\bea
(F\Rightarrow G)(p)&=&\{
	r\in\C :\mbox{for any causal relation }f:p\rightarrow q,\nonumber\\
	& &\ \mbox{if }\ r\in F(q), \mbox{ then } r\in G(q)\}.
\eea
That is, $(F\Rightarrow G)(p)$  contains all those events, which,
if by the time step $f:p\rightarrow q$ join $F$ at $q$, they also join 
$G$ at $q$. One can check that this is a ${\bf Set}^\C$ functor. 
\item
Complement:\\
Since we have already worked with the complement of $\Past$, 
we will discuss the complement using this particular example
rather than the abstract functor $F$. We will basically 
give the justification for the definition (\ref{eq:negpast}).
The problem with applying the definition (\ref{eq:setcomplement}) 
of a boolean complement 
to $\Past$ is that it gives its complement as
all $r\in\C$ that are not in $\Past(p)$, namely
$(\C-\Past(p))$. But, since a causal past keeps
getting larger with time, i.e.\   $\Past(p)\subseteq
\Past(q)$ when $p\leq q$, this complement would get smaller in time, 
$(\C-\Past(p))\supseteq (\C-\Past(q))$. This has the wrong behavior to be
a functor like $\Past$ and hence we cannot get an 
algebra of evolving sets using this complement.  

Let us instead define the complement by generalising 
(\ref{eq:setcompfromimpl}). First we need an evolving 
version of $\emptyset$. This is the functor $\Empty$
that assigns to each $p\in\C$ the empty set (with the expected 
inbetween arrows). We then get
\beq
\neg \Past=(\Past\Rightarrow \Empty).
\label{eq:negfromimply}
\eeq
In components this gives
\bea
\neg \Past(p)&:=&\{r\in\C:\mbox{for any causal relation }f:p\rightarrow q
			\mbox{ in }\C,\nonumber\\
		& &	\ f(r)\notin \Past(q)\}\nonumber\\
		&=&\{r\in\C:\ r\notin \Past(q)\mbox{ for any } q\geq p\}.
\label{eq:pastcomplement}
\eea
which is the definition (\ref{eq:negpast}).
This is a ${\bf Set}^\C$ functor and an appropriate complement 
for $\Past$. 
\end{itemize}

Generally, $\neg \Past$ satisfies 
\beq
\Past\cap\neg \Past= \Empty,
\label{eq:HAneg1}
\eeq
but 
\beq 
\Past\cup\neg \Past\subset \World.
\label{eq:HAneg2}
\eeq
(An example of this is the case of a lattice causal set where the 
righthand side of (\ref{eq:HAneg2}) is just $\Past$.)
Because the Heyting algebra complement satisfies the weaker
set of conditions (\ref{eq:HAneg1}) and (\ref{eq:HAneg2}),
instead of (\ref{eq:setcomplement}), it 
is often called a {\it pseudo-complement}. 
We simply call it complement,
or a Heyting algebra complement. For a Heyting algebra complement,
it is not true anymore that $\neg\neg \Past$ equals $\Past$
(again consider the example where $\C$ is a lattice).   
In general, for some evolving set, {\it the double complement is 
larger than the original}.

The subobjects of a ${\bf Set}^\C$ object like $\World$, 
therefore, form a lattice: there is union and intersection 
of any two, $\World$ is the supremum and $\Empty$ the infimum. 
This is independent of whether the causal set itself is a lattice. 
This is significant because a lattice can be used as an algebra
and, at the same time, evolving sets 
still contain all the causal information in $\C$.

The lattice of evolving sets
is a distributive one, since for any three subobjects 
$F,G$ and $H$ of $\World$,
\beq
F\cap(G\cup H)=(F\cap G)\cup(F\cap H)
\eeq
holds. 

A lattice   that is distributive and has
implication defined for every pair of elements is called 
a {\it Heyting algebra}. It is weaker than the boolean algebra 
(it is a generalisation).
For a Heyting algebra element (evolving set) $F$,
\beq
\neg \neg F\supseteq F.
\eeq
(but one can check that $\neg\neg\neg F=\neg F$).
If it is the case that $\neg F\cup F=\World$, then
$\neg \neg F=F$ and  the Heyting algebra reduces to a boolean
one.

Summarising, the algebra of the 
analogue of the powerset of an evolving set
is a Heyting algebra.
A key feature of a Heyting algebra is the complement. 
It is intimately tied to the causal structure 
and therefore has physical significance.

\section{Using evolving sets in a causal quantum theory}
\label{quantum}

The reason we carried out the analysis of this paper is that 
it may serve in obtaining a quantum theory of gravity.
We are keeping as much of the causal structure of general 
relativity as it is possible without using metric manifolds. 
Clearly, we are restricted to very basic features of a spacetime,
namely, the causal ordering of events. No further qualifications
of these events have been specified. It is true that it is 
possible to recover the metric of a spacetime up to a conformal 
factor from the causal relations
between all its events. It has been argued that in the discrete case,
even this factor may be fixed by assigning spacetime volume to 
each event \cite{sorkinvol}. However, in the causal set as it is used here,
no straightforward relationship between its events and those of a 
classical spacetime has been assumed. In this case,
the causal structure is not all that is needed. 

As it has been proposed in \cite{fmls1} and further elaborated in 
\cite{dual,fmls2,BG}, extra spatial structure can be attached to
a  causal set in the form of a spin network, a graph with edges 
labelled by representations of $SU(2)$ \cite{penrose}. 
Spin networks have already been
part of a quantized gravity theory in loop quantum gravity \cite{spinnets}. 
On the other hand, in the causal spin network scheme,
as in other applications of causal sets, a problem is how to represent
the causal structure in a way that is useful in the quantum theory.
We hope that the present algebraic formulation of causality will 
resolve this, at least by restricting the quantum theory 
to the functorial form that implies underlying causality,
as we will outline below. The application of evolving sets to spin
networks and generally a causal quantum theory is currently being
constructed. Here we indicate how this application may be carried out. 

\subsection{Elementary causal quantum theory}
\label{hilb}

Standard quantum theory is unambiguously applicable under 
certain conditions: that the time is fixed and that the 
separation between the system in question and its environment 
is also fixed.\footnote{
At the conceptual level, varying sets is the method proposed by 
Isham \cite{cji} to accomodate contextual physical statements, where
context means the conditions under which a certain physical description
and physical statement is applicable. We have dealt with evolving 
sets (and in this section
we discuss what may be called ``evolving algebras'' and 
``evolving state spaces'') where the context is causality, namely
we need to specify the event $p$ in order to get its 
classical past or the quantum theory at $p$.  
The same method can be used in a third context, to treat the 
system-environment split implied by standard quantum theory.
Here the context is all possible systems in the universe. This 
is work in progress with C. Rovelli.}
Quantum field theory  is, in principle, more flexible, however, the 
single Hilbert space employed restricts its use to a 
semiclassical regime of quantum gravity. 

Leaving aside for the time being the system/environment issue,
and keeping the causal set as the time evolution record,
we may propose removing the fixed-time condition by using a 
Hilbert space for each event in the causal set.
For example, in the Newtonian case, this results in a sequence of 
Hilbert spaces which we may regard as the evolution of the 
quantum universe. 

Now we note the following. 
The present work suggests that the Hilbert spaces at each event 
may be regarded as objects in the category of Hilbert spaces,
${\bf Hilb}$ (with linear transformations as the arrows).
Then, the causal structure is preserved in the quantum theory if 
there is a functor from the causal set to ${\bf Hilb}$,
$Q:\C\rightarrow{\bf Hilb}$. Let us check if this is the case. 
For each event $p\in\C$, we have a Hilbert space $H(p)=Q(p)$.
However, given a causal relation $p\leq q$ in $\C$, 
it is not, in general, the case that there is a linear map 
from $H(p)$ to $H(q)$ such that $H(p)$ is a linear subspace 
of $H(q)$. Therefore,  the ordering of $\C$ is not preserved 
by the functor $Q$ into ${\bf Hilb}$. 

Some of the ways to circumvent this problem and maintain causality in the 
quantum theory that we are aware of, is to restrict the linear 
maps, evolve linear subspaces of the above Hilbert spaces (evolve
the projection operators), and replace the Hilbert spaces with 
algebras of observables at each event (see \ref{obsquantum}).  

\subsection{Causal spin networks}
\label{snetquantum}

Causal evolution of spin networks is in principle similar to the 
above scheme, but with the important difference that 
it leads to entangled states rather than states in a single
Hilbert space (one Hilbert space for each event). 
Analyzing a functor from the causal set to entangled states or
density matrices is 
beyond the scope of this paper, but we will illustrate how they arise. 

One may regard a causal history of spin networks as a 
sequence of spin network graphs such that the nodes of all the graphs 
in a given history are
the elements of the causal set and therefore are partially 
ordered.\footnote{In fact, it is the local changes of graphs which 
generate the history that are partially ordered into a causal set. 
The difference does not affect the present discussion. For more
detail on this, see \cite{fmls2}.}
Between the nodes of a single spin network, therefore, there are no
causal relations. Assuming a causal history with connected graphs,
the set of nodes of a spin network is maximal, i.e.\ there is not 
an element in $\C$ outside this set that is not causally related to 
some element in the set. A maximal set of causally unrelated events 
in $\C$ is a {\it maximal antichain}.

The complaint with which we started this paper, that a spatial slice 
can only be seen by an observer in the infinite future, is also 
present in the spin network evolution since an antichain requires the
same observer. 
What this work suggests is that we should instead work with 
{\it (partial) antichains at event $p$},
for each $p\in\C$ ($p$ is also a spin network node).
These are sets of nodes in a causal 
spin network history that are causally unrelated {\it inside}
$\Past(p)$. 

Given the functor $\Past$, we can construct the functor
$\mbox{\sl Antichains}:\C\rightarrow{\bf Set}$,
\beq
\mbox{\sl Antichains }(p)=
		\{\mbox{sets of causally unrelated events in }\Past(p)\}.
\eeq
For every causal relation $p\leq q$, the antichains at $p$ are,
of course, a subset of those in $q$.
 
Given an element in $\mbox{\sl Antichains }(p)$,
there is a number of graphs that has these events as its nodes, i.e.,
for each $p$ we have the set $Graphs(p)$ of all graphs with 
elements of $\mbox{\sl Antichains}(p)$ as their 
sets of nodes. 
(We ignore the $SU(2)$ labels which further enlarge this set.)
Clearly, $Graphs(p)$ includes open spin networks. 

An open spin network  has  free edges, all labelled by  
$SU(2)$ representations, and thus can be regarded as an entangled state. 
Hence, to go from open spin networks at 
$p$ to open spin networks at a later $q$, we need an evolution 
operator on density matrices. 

\subsection{Observable causal quantum theory}
\label{obsquantum}

A further possibility is 
to replace the Hilbert spaces of the scheme in \ref{hilb}
with the algebra of observables at each event.\footnote{
I need to thank Eli Hawkins for this observation.}
This means using the functor
\beq
\sl{OQ}:\C\longrightarrow {\cal A}
\eeq
where ${\cal A}(p)$ will be  the algebra of observables at $p$.

This scheme provides what we may regard as a quantum field theory on 
a causal set.

\vskip 0.8cm
Concluding this section, we note that the use of functors from 
the causal set to our preferred description of the universe
is a safe way to check that this description is indeed causal.
If the quantum theory is the output of a functor that has the causal 
set as its domain, then causality is build into the theory.
Even if causality is not the preferred set of conditions on 
evolution (one may wish to impose a weaker ordering 
than the causal set, for example a non-transitive order), 
the same method can be used to check that 
these conditions are preserved in the quantum theory.

\section*{Conclusions}

In this paper, we used the functor $Past:\C\rightarrow {\bf Set}$
to transform from an external (outside the universe) viewpoint of
causality to the internal, finite-time, viewpoint provided by the
components of $Past$ at each event.
The result provides an algebraic description of causality 
and great possibilities for generalisations to diverse types of
causal quantum theories, some of which we outlined above. 
Conclusions about the form of quantum evolution that causality permits 
can be reached in this way. However, note that this
would be considered a kinematical restriction 
since the causal set on which the evolving set is defined is 
fixed.

$\Past$ is the simplest evolving set, since it only uses the
causal set and no additional spatial or field information.  
It does, however, serve as 
a good example in introducing evolving sets, the Heyting 
algebra of evolving powersets, and the non-standard evolving
complement. At the classical level (from the perspective of a 
causal set approach), the technical advantage of 
$\Past$ is that it transforms
the partially ordered set of causal relations into a lattice
and suggests significant improvements in the description of 
the causal structure without the use of a spacetime.\footnote{
After the work in this paper was completed, the paper of Bombelli and
Meyer \cite{BM} came to our attention. From a given causal 
set, they construct quantities that 
contain the same events as $\Past$ and $\neg \Past$.}
The actual implementation in the quantum theory 
will be reported in future work. 

Further, it is possible to construct a framework for a (classical) 
causal cosmological theory by requiring that every physical 
observable corresponds to an observation made by an observer inside 
the universe, represented by an event or a 
collection of events in the set of causal relations of that 
universe.   Such ``internal'' observables, when referring to events 
in the observer's causal past, are subfunctors of $\Past$.  
For example, in the $1+1$ causal histories of Abjorn, Loll and others 
\cite{AL}, a physical observable may be events that have 
spacetime valence $n$ (the number of ingoing and outgoing causal 
relations to such an event is $n$).  Written as a varying set, this 
observable will be the $n$-valent events that have occured at $p$ and 
is a subfunctor of $\Past$.  Such internal causal observables obey a 
Heyting algebra. 

We close by noting that sieves can be used to specify time. This is an 
algebraic alternative to time expressed as $t\in{\bf R}$.  In work 
currently in progress, we investigate projection 
operators which are time-dependent in the sense that their 
eigenvalues belong to a larger set than the standard $1$ and $0$; 
in fact they correspond to sieves. 

\section*{Acknowledgments}

I am very grateful to Chris Isham for introductory discussions
on varying sets and his very useful suggestions and criticisms 
on the present work. Detailed comments on the first draft from 
Eli Hawkins, Carlo Rovelli, Lee Smolin and Adam Ritz have made this
paper much clearer than it originally was. I am thankful to 
Sameer Gupta, David Meyer and Roger Penrose for suggestions on using 
the complement to define a black hole. Thanks are due to John 
Baez, John Barrett and Alex Heller and particularly
to Bas van Fraasen for the first discussions on intuitionistic logic.  

This work was supported by NSF grants PHY/9514240 and PHY/9423950
to the Pennsylvania State University and a gift from the Jesse 
Phillips Foundation.

\section*{Appendix: Boolean vs.\ intuitionistic logic}
\label{logic}

This appendix is an account of the basics of Boolean and 
intuitionistic logic and their relation to the Boolean and 
Heyting algebras.

The boolean algebra obeyed by sets means
that when a physical theory is ultimately built on 
set-theoretic foundations
(which is almost universally the case), the underlying logic is 
boolean. Loosely speaking, an observer in such a theory will 
make statements which obey boolean logic. For 
a theory based on evolving sets, which we propose here, 
the Heyting algebra of evolving sets indicates that the 
underlying logic is {\it intuitionistic}. 
For completeness, and because we would like to 
point the physicist reader to a mathematical literature possibly
of use in issues of time evolution in physics, this appendix is 
a review of both boolean and intuitionistic logic.

From the perspective of a physicist, 
there is little difference between the algebraic operations and
the corresponding logical ones (propositional calculus). 
Practically, it simply 
involves ``reading'' the operations  as propositions rather 
than sets. For example, each proposition $x$ in the logical 
operations that follow may be replaced by ``$a\in A$'' for some subset 
$A$ of $X$ which we used in section \ref{algebra}.
The algebraic operations of union, intersection, implication,
and complement then become the logical connectives
{\small OR, AND, IMPLIES} and {\small NOT}.

Most of the interpretational discussion has already been 
carried out in section \ref{algebra}. We now simply present the 
two logical systems. In the intuitionistic case, we include in the
presentation some of the historical reasons for the introduction
of this type of logic into mathematics.

\subsection{Boolean logic}

To have either an algebra or a propositional calculus we 
first need a lattice, as explained in section \ref{algebra}.
A lattice is first of all a partial order, so we partial order 
propositions by $\leq$. The proposition $x\leq y$ is to be read as
``if $x$ is true, then $y$ is true''.

For a pair of propositions $x$ and $y$, the four boolean logical 
operations {\small OR, AND, IMPLIES} and {\small NOT}
produce four new propositions. Any further statements made
(sentences) are constructed by combinations of these
basic operations. They are in obvious correspondence 
with the four boolean algebra operations.  

\begin{itemize}
\item
{\small OR} (union): $(x\vee y)$ is true when 
either $x$ or $y$ are true.
\item
{\small AND} (intersection): $(x\wedge y)$ is true
when both $x$ and $y$ are true. 
\item
{\small IMPLIES} (implication):  $(x\Rightarrow y)$ is true when, 
if $x$ is true, then $y$ is also true. 
\item
{\small NOT} (complement): The proposition $\neg x$ is true
whenever $x$ is false. 
\end{itemize}

\noindent We can tabulate the boolean operations:
\begin{tabular}{|c|c||c|c|c|c|}\hline
$x$ 	& $y$ 
	& $x\vee y$ & $x\wedge y$& $x\Rightarrow y$ &$\neg x$\\ \hline\hline
1	&1	&1	&1	&1	&0 \\ \hline
1	&0	&1	&0	&0	&0 \\ \hline
0	&1	&1	&0	&1	&1 \\ \hline
0	&0	&0	&0	&1	&1 \\ \hline
\end{tabular}
\vskip 0.2cm
Note, first, that $\Rightarrow$ as defined above means that 
\beq
(x\Rightarrow y)=\neg x\vee y,
\eeq
that is ``$x$ implies $y$'' is equivalent to ``either $x$ is 
false or $y$ is true'', which with little thinking we can believe
(or use the table above).
 
Second, in a boolean algebra, every proposition $x$ has a negation $\neg x$
which satisfies
\beq
x\wedge \neg x=0\qquad\mbox{and}\qquad x\vee\neg x=1,
\eeq 
where 0 means false and 1 means true.
As a result, not not $x$ is the same as $x$,
\beq
\neg\neg x=x.
\eeq

\subsection{Intuitionistic logic}

We think it is useful to present intuitionistic logic 
with reference to the motivation of the mathematicians who invented it. 
This is standard material and we 
freely quote from the introduction of \cite{McLM} and   from 
\cite{constructive}.

Intuitionistic logic and the mathematics based on it 
originated with Brouwer's work on the foundations of mathematics at
the beginning of this century \cite{brouwer}. He  
insisted that all proofs be constructive. This means that 
he did not allow proof by contradiction and hence he excluded the
classical (boolean) ``for all $x$, either $x$, or not $x$''. 
Intuitionism is a form of constructive mathematics.
The classical (boolean) mathematician believes that every mathematical 
statement $x$ is true or false, whether or not he has the proof for it. 
The constructive mathematician does not consider $x$ to be true or 
false unless he can either prove it or disprove it. That is, 
$x$ may be true tomorrow, or false tomorrow. 

To quote Brouwer, ``the belief in the universal 
validity of the excluded middle in mathematics is considered by the 
intuitionists as a phenomenon in the history of civilization of
the same kind as the former belief in the rationality of $\pi$,
or the rotation of the firmament about the earth''.
 
Brouwer's  approach was not 
formal or axiomatic, but subsequently Heyting and others 
introduced formal systems of intuitionistic logic, weaker than classical 
logic.  Heyting first formalised the basic axioms of  intuitionism 
which, usually detached from Brouwer's extreme position,
has turned out to be useful for mathematics beyond the original context.

As in the Boolean case, we start with a lattice of propositions, 
ordered by $\leq$. 
The four operations work as follows:
\begin{itemize}
\item
{\small OR} and {\small AND} are the same as in the boolean case. 
\item
{\small IMPLIES}:
The statement $(x\Rightarrow y)$ means that $y$ holds under the assumption 
that $x$ holds, namely, we show $(x\Rightarrow y)$ by deriving 
$y$ from the hypothesis $x$. (Note that there is no causal implication 
in $\Rightarrow$, there is no sense of $y$ causally following $x$.)
$(x\Rightarrow y)$ is characterized by
\beq
z\leq(x\Rightarrow y)\qquad\mbox{if and only if}\qquad
	z\wedge x\leq y,
\label{eq:logicimply}
\eeq
that is, $(x\Rightarrow y)$, namely the condition (\ref{eq:logicimply}),
 is the union of all $z$ that satisfy 
$z\wedge x\leq y$.\footnote{
In the open set example (footnote 8), $(U\Rightarrow V)$ is the union of all 
open sets $W_i$ whose intersection with $U$ is included in $V$,
i.e.\ $(U\Rightarrow V)=\bigcup_i W_i$ where $W_i\cap U\subset V$.}

Note that the constructive interpretation of $(x\Rightarrow y)$ is 
weaker than the boolean one where $(x\Rightarrow y)=\neg x\vee y$.
This leads to the modified intuitionistic negation.

\item
{\small NOT}:
The statement $\neg x$ means that $(x\Rightarrow z)$, where $z$ is a 
contradiction. Usually a contradiction is denoted $0$, so
\beq
\neg x= (x\Rightarrow 0).
\eeq
that is, we have ``not $x$'' when $x$ leads to a contradiction. 
\end{itemize}
 
From the definition  (\ref{eq:logicimply}) of $\Rightarrow$, we get that
\beq
y\leq \neg x\qquad\mbox{if and only if}\qquad y\wedge x=0,
\eeq
namely, $\neg x$ is the union of all propositions $y$ which have nothing 
in common with $x$. As a result $\neg\neg x$ needs not equal $x$. 
Also, although $x\wedge\neg x=0$, it may not be the case that $x\vee 
\neg x=1$. 

In short, intuitionistic logic, or, equivalently, a Heyting algebra, 
is particularly
suitable in a theory with time evolution, when we are concerned with 
physical statements which become true at a certain time stage and 
stay true afterwards.

\end{document}